\documentclass[a4paper,12pt, oneside]{article}
\usepackage[dvipdfm]{graphicx}% Include figure files
\usepackage{dcolumn}% Align table columns on decimal point
\usepackage{bm}% bold math
\begin{document}
\baselineskip 24pt
%\begin{spacing}{2}

\begin{center}
{\Large
X-ray absorption spectroscopy and X-ray magnetic circular dichroism studies of transition-metal-co-doped ZnO nano-particles}
\vspace{1cm}

\rm{T.~Kataoka, Y.~Yamazaki, V.~R.~Singh, Y.~Sakamoto, K.~Ishigami,\\ V. K. Verma, and A. Fujimori}\\
\it{
Department of Physics and Department of Complexity Science and Engineering,
University of Tokyo, Bunkyo-ku, Tokyo 113-0033, Japan
}%
\vspace{0.5cm}

\rm{F.-H.~Chang, H.-J.~Lin, D.~J.~Huang, and C.~T.~Chen}\\
\it{
National Synchrotron Radiation Research Center, Hsinchu 30076, Taiwan
}%
\vspace{0.5cm}
%%%%%%%%%%%%%%%%%%%%%%%%%%%%%%%%%%%%%%%%%%%%%%%%%%%%%%%%%%%%%%%%%%%%%%%%%

\rm{D.~Asakura and T.~Koide}\\
\it{
Photon Factory, IMSS, High Energy Accelerator Research Organization, Tsukuba, Ibaraki 305-0801, Japan
}%
\vspace{0.5cm}
%%%%%%%%%%%%%%%%%%%%%%%%%%%%%%%%%%%%%%%%%%%%%%%%%%%%%%%%%%%%%%%%%%%%%%%%%%%%%

\rm{A.~Tanaka}\\
\it{
Department of Quantum Matter, ADSM, Hiroshima University, Higashi-Hiroshima 739-8530, Japan
}%
\vspace{0.5cm}
%%%%%%%%%%%%%%%%%%%%%%%%%%%%%%%%%%%%%%%%%%%%%%%%%%%%%%%%%%%%%%%%%%%%%%%%%%%%%%%

\rm{D.~Karmakar}\\
\it{Technical Physics Division, Bhabha Atomic Research Center, Mumbai 400085, India
}%
\vspace{0.5cm}

\rm{S.~K.~Mandal and T.~K.~Nath}\\
\it{Department of Physics and Meteorology, Indian Institute of Technology, Kharagpur 721302, India
}%
\vspace{0.5cm}

\rm{I.~Dasgupta}\\
\it{Department of Solid State Physics Indian Association for the Cultivation of Science Jadavpur, Kolkata 700 032, India
}%
\vspace{0.5cm}

\end{center}

\date{\today}

\newpage
\begin{center}
\section*{Abstract}
\end{center}
We report on x-ray absorption spectroscopy (XAS) and x-ray magnetic circular dichroism (XMCD) studies of the
paramagnetic (Mn,Co)-co-doped ZnO and ferromagnetic (Fe,Co)-co-doped ZnO nano-particles.
Both the surface-sensitive total-electron-yield mode and the bulk-sensitive total-fluorescence-yield mode have been employed to extract
the valence and spin states of the surface and inner core regions of the nano-particles.
XAS spectra reveal that significant part of the doped Mn and Co atoms are found in the trivalent and tetravalent state in particular in the surface region
while majority of Fe atoms are found in the trivalent state both in the inner core region and surface region.
The XMCD spectra show that the Fe$^{3+}$ ions in the surface region give rise to the ferromagnetism
while both the Co and Mn ions in the surface region show only paramagnetic behaviors.
The transition-metal atoms in the inner core region do not show magnetic signals, meaning that they are antiferromagnetically coupled.
The present result combined with the previous results on transition-metal-doped ZnO nano-particles and nano-wires suggest
that doped holes, probably due to Zn vacancy formation at the surfaces of the nano-particles and nano-wires,
rather than doped electrons are involved in the occurrence of ferromagnetism in these systems.

\newpage

%%%%%%%%%%%%%%%%%%%%%%
\section{Introduction}
%%%%%%%%%%%%%%%%%%%%%%

Various semiconducting oxides such as ZnO \cite{Sharma}, TiO$_2$ \cite{Yamasaki}, and SnO$_2$\cite{Wang} in thin film and
nano-particle forms are known to exhibit ferromagnetism at room temperature when they are doped with transition-metal atoms.
Current interest in such magnetic nano-particle systems is motivated by unique electronic structures and
magnetism at the surfaces of the nano-particles which are different from the inner core region.
In the nano-particle form, the structural and electronic properties are modified by
surface defects such as Zn and O vacancies with broken chemical bonds and charge imbalance,
which may mediate or modify exchange coupling between the doped atoms \cite{NGanguli}.
For example, in the case of (Mn,Co)-co-doped ZnO [ZnO:(Mn,Co)] nano-particles \cite{KataokaMnCo},
high-valence (3+ and 4+) Mn and Co ions are found to be present, probably due to the formation of Zn vacancies (V$\rm_{Zn}$) in the surface region.
The doped Fe atoms in the ferromagnetic ZnO nano-particles are converted from 2+ to 3+
due to hole doping in the surface regions \cite{NGanguli, KataokaFe, KarmakarFe}, resulting in the ferromagnetic interaction between the doped Fe atoms.
In the case of Co-doped ZnO systems such as (Co,Ga)-co-doped ZnO \cite{YHe} and Co-doped ZnO nano-particles \cite{HGu},
on the other hand, oxygen vacancies (V$\rm_{O}$), which induce electron doping, are reported to be necessary for ferromagnetism.
Recently, room-temperature ferromagnetism was reported for (Fe,Co)-co-doped ZnO [ZnO:(Fe,Co)] in thin film \cite{YMCho}
and nano-particle forms \cite{KarmakarFeCo}.
From the first-principle calculations, Karmakar ${et}$ ${al}$. \cite{KarmakarFeCo} have indicated
that V$\rm_{Zn}$-mediated double exchange interaction plays important role for ferromagnetism in ZnO:(Fe,Co) nano-particles.
Indeed, enhancement of ferromagnetic interaction between transition-metal atoms has been demonstrated in
previous first-principles calculations by Gopal and Spaldin \cite{Gopal}.
First-principle calculations by Park and Min \cite{MSPark}, on the other hand, have suggested the importance of
RKKY-type exchange interaction mediated by conduction carriers induced by V$\rm_O$ as the origin of ferromagnetism of ZnO:(Fe,Co).
Also, calculations by Ghosh ${et}$ ${al}$. \cite{SGhosh} have indicated direct exchange interaction
mediated by the doped electron carriers at the Fe-V$\rm_O$-Co defect configuration in the surface region of ZnO:(Fe,Co) nano-wires.

Thus, it has been controversial whether the enhancement of exchange interaction comes from electron doping or hole doping.
In this paper, we report on x-ray absorption spectroscopy (XAS) and x-ray magnetic circular dichroism (XMCD) studies of
paramagnetic ZnO:(Mn,Co) and ferromagnetic ZnO:(Fe,Co) nano-particles.
The valence and spin states of the doped ions and their magnetic interaction have been revealed by XAS and
XCMD measurements of the transition-metal core levels.
Also, both the surface-sensitive total-electron-yield mode and the bulk-sensitive total-fluorescence-yield mode have been employed to extract
the valence and spin states of the surface and inner core regions of the nano-particles separately.
The experimental results indicate that doped holes rather than doped electrons are involved in the occurrence of ferromagnetism in these systems.

%%%%%%%%%%%%%%%%%%%%%%
\section{Experimental Methods}
%%%%%%%%%%%%%%%%%%%%%%
Transition-metal-co-doped ZnO nano-particles were synthesized by a low temperature chemical pyrophoric reaction process.
We have prepared paramagnetic ZnO:(Mn,Co) nano-particles (Mn=15 \%, Co=15\%), and
ferromagnetic ZnO:(Fe,Co) nano-particles (Fe=5 \%, Co=5\%) with $T_C$ $>$ 300 K.
Details of the sample preparation were described in refs.\cite{KarmakarFeCo, KarmakarFe, SKMandal}.
Structure characterization was carried out by x-ray diffraction (XRD), selected area electron diffraction (SAED) and
transmission electron microscopy (TEM).
We have made pellets from calcined powders and then sintered them at a temperature of $\sim$ 570 K for 30 min.
The average size of the nano-particles were 7-10 nm \cite{KarmakarFeCo, KarmakarFe}.

XAS and XMCD measurements of ZnO:(Fe,Co) samples and XAS measurements of ZnO:(Mn,Co) samples
were performed at the Dragon Beamline BL-11A of National
Synchrotron Radiation Research Center (NSRRC), Taiwan.
The spectra were taken both in the total-electron-yield (TEY: probing depth $\sim$ 5 nm) and the
total-fluorescene-yield (TFY: probing depth $\sim$ 100 nm) modes, i.e.,
the TEY and TFY modes are relatively surface- and bulk-sensitive, respectively.
The degree of circular polarization of x-rays was $\sim$ 60\%.
XAS and XMCD measurements of ZnO:(Mn,Co) samples were also made at BL-16A of Photon Factory (KEK-PF).
The degree of circular polarization of x-rays was more than $\sim$ 90\%.
All the measurements were performed at room temperature.

Absorption spectra were analyzed using configuration-interaction (CI) cluster-model calculations.
The cluster consisted of a transition-metal ion octahedrally and/or tetrahedarally coordinated by O$^{2-}$ ions.
The ground state wave function was expanded in the
$\psi$= $\alpha$$|d^n\rangle$ + $\beta$$|d^{n+1}\underline{L}\rangle$ + $\gamma$$|d^{n+2}\underline{L^2}\rangle$,
where $\underline{L}$ denotes an ligand O 2$p$ hole.
The adjustable parameters of the calculation were the charge-transfer energy $\Delta$,
the $d$-$d$ Coulomb energy $U$, the $p$-$d$ transfer integral $T$, and the crystal field splitting parameters 10Dq.
We assumed high-spin states for the calculations, and 10Dq was assumed to be less than 1.0 eV.

\section{Results and Discussion}

Figures 1(a) and 1(b) show the Mn and Co 2$p$$\rightarrow$3$d$ XAS spectra of the paramagnetic ZnO:(Mn,Co) nano-particles,
respectively, taken both in the TEY and TFY modes.
In the figures, we compare the experimental spectra (circles) taken both in the TEY and TFY modes with the cluster-model calculations for the Mn and Co ions
with various valence states, tetrahedrally co-ordinated by oxygen atoms \cite{ATanaka}.
From the line-shape analysis shown in Figs. 1(a) and 1(b),
the relative concentrations of Mn$^{2+}$ and Co$^{2+}$ ions estimated using TFY mode are higher than those estimated using TEY mode
because the features due to the Mn$^{2+}$ and Co$^{2+}$ states in the TEY mode are weak compared to those in the TFY mode.
This indicates that the relative concentrations of Mn$^{2+}$ and Co$^{2+}$ ions are relatively high in the inner core region of the nano-particles
and those of the higher valence states of Mn$^{3+}$, Mn$^{4+}$, Co$^{3+}$, and Co$^{4+}$ are relatively high in the surface region.

Figures 2(a) and 2(b) show the Mn and Co 2$p$$\rightarrow$3$d$ XAS and XMCD spectra of the paramagnetic ZnO:(Mn,Co) nano-particles, respectively, taken in the TEY mode.
We compare the Mn 2$p$$\rightarrow$3$d$ XMCD spectra of  the ZnO:(Mn,Co) nano-particles
with those of Ca$_{1-x}$Mn$_x$RuO (CMRO) \cite{Terai} and Zn$_{1-x}$Mn$_x$Se$_2$ \cite{Hofmann}, and
compare the Co 2$p$$\rightarrow$3$d$ XMCD spectra of  the ZnO:(Mn,Co) nano-particles with that of Ti$_{1-x}$Co$_x$O$_2$\cite{Mamiya}.
It is likely that Mn 2$p$$\rightarrow$3$d$ XMCD spectrum comes from the Mn$^{3+}$ and Mn$^{4+}$ ions
because the line shape of XMCD spectrum of ZnO:(Mn,Co) is similar to that of CMRO, where Mn$^{3+}$ and Mn$^{4+}$ ions coexist.
The Co 2$p$$\rightarrow$3$d$ XMCD spectral line shape of the ZnO:(Mn,Co) nano-particles is similar to that of Ti$_{1-x}$Co$_x$O$_2$.
From the experimental results, we suggest that paramagnetic component of the XMCD signals
consists of the Mn$^{3+}$, Mn$^{4+}$ and Co$^{2+}$ states.

Figures~3(a) and~3(b) show the Fe and Co 2$p$$\rightarrow$3$d$ XAS spectra of the ferromagnetic ZnO:(Fe,Co) nano-particles, respectively.
In the figures, we compare the experimental spectra (circles) taken both in the TEY and TFY modes with the cluster-model calculations for the Fe and Co ions
with various valence states, tetrahedrally or octahedrally co-ordinated by oxygen atoms \cite{ATanaka}.
In the transition-metal-doped ZnO nano-particles, the valence and the co-ordination of
the doped atoms will be 2+($T_d$) if no vacancies are created, or often become 3+($T_d$) or 3+($O_h$) due to the vacancy formation in the surfaces \cite{NGanguli, KataokaFe}.
We therefore calculated spectra for the 2+($T_d$), 3+($T_d$), and 3+($O_h$) states of the Fe and Co ions.
Here, $O_h$ is an interstitial site of the Wurzite-type ZnO lattice.
From the line-shape analysis shown in Fig. 3(a), one notices that the Fe ions in the surface region are mostly Fe$^{3+}$($O_h$) with a small amount of Fe$^{2+}$($T_d$).
In the experimental XAS spectra taken in the TFY mode, the dip structure at 710 eV is shallower, that is,
the Fe$^{2+}$($T_d$) component increases in the inner core region, suggesting that Fe$^{3+}$($O_h$) ions mainly come from the surfaces.
From the Co 2$p$$\rightarrow$3$d$ XAS spectra, it is likely that the doped Co atoms in the surface region are Co$^{2+}$($T_d$), Co$^{3+}$($T_d$) and Co$^{3+}$($O_h$).
On the other hand, the Co atoms in the inner core region appear to be largely in the Co$^{2+}$($T_d$) state.

Figures 4(a) and 4(b) show the Fe 2$p$$\rightarrow$3$d$ XAS and XMCD spectra of the ferromagnetic ZnO:(Fe,Co) nano-particles, respectively, taken at $H$$=$1 T.
The Fe 2$p$$\rightarrow$3$d$ XMCD intensity taken in the TEY mode was finite,
while the XMCD spectrum taken in the TFY mode showed low intensity and not clear observed.
This indicates that the Fe ions in the surface region but not in the inner core region are magnetically active.
Also, one notices that XMCD signals at the Fe $L_2$ absorption edge are very weak,
suggesting that a large orbital magnetic moment ($M_{orb}$) of the Fe ion, probably due to a mixture of Fe$^{2+}$ component.
In the nano-particle form, which has a relatively large surface area, the spin-orbit coupling and
magnetic anisotropy may be enhanced due to surface effects.
Indeed, this large $M_{orb}$ has been observed for ZnO:Fe nano-particles \cite{KataokaFe}.
Figures 4(c) shows the Fe 2$p$$\rightarrow$3$d$ XMCD spectra taken in the TEY mode  at various magnetic fields.
In Fig. 4(d), the XMCD intensities due to Fe$^{3+}$($O_h$) and Fe$^{2+}$($T_d$) are plotted as a function of magnetic field.
The intensity due to Fe$^{3+}$($O_h$) increases with magnetic field but persists at low fields down to $H$$=$0.2 T,
while the XMCD intensity due to Fe$^{2+}$($T_d$) remains unchanged with magnetic field.
These results indicate that Fe$^{3+}$($O_h$) contributes to both the ferromagnetism and paramagnetism and
that Fe$^{2+}$($T_d$) contributes only to the ferromagnetism.

Figures 5(a) and 5(b) show the Co 2$p$$\rightarrow$3$d$ XAS and XMCD spectra of the ferromagnetic ZnO:(Fe,Co) nano-particles, respectively, taken at $H$$=$1 T.
The Co 2$p$$\rightarrow$3$d$ XMCD intensity taken in the TEY mode was finite,
while the XMCD intensity taken in the TFY mode did not show finite intensity.
This suggests that the Co ions in the surface region are magnetically active as in the case of Fe.
One can see that the Co 2$p$$\rightarrow$3$d$ XMCD spectrum, taken in the TEY mode, comes from the Co$^{2+}$($T_d$) and Co$^{3+}$($T_d$) ions.
Figures 5(c) shows the Co 2$p$$\rightarrow$3$d$ XMCD spectra taken at various magnetic fields,
and Fig. 5(d) shown the Co 2$p$$\rightarrow$3$d$ XMCD intensity as a function of magnetic field.
This increases with magnetic field, indicating that the ionic Co atoms in the surface region is paramagnetic
and that the ferromagnetic component of the Co ions is negligibly small.
The negligibly weak XMCD signals in the spectra recorded in the TFY mode indicate that the Co ions in the inner core region is antiferromagnetically coupled,
We thus conclude that the ferromagnetism of the ZnO:(Fe,Co) nano-particles comes only from the Fe ions in the surface region.

It should be noted that the Fe 2$p$$\rightarrow$3$d$ XMCD spectra of ZnO:(Fe,Co) indicate the spins of
Fe$^{3+}$($O_h$) and Fe$^{2+}$($T_d$) signals to be in the same directions.
Therefore the segregation of  ferromagnetic or ferrimagnetic Fe oxides such as  ZnFe$_2$O$_4$ \cite{SNakashima, MHofmann},
$\gamma$-Fe$_2$O$_3$ \cite{SBProfeta}, and Fe$_3$O$_4$ \cite{RCornell} can be excluded because
in these materials Fe$^{3+}$($T_d$) and Fe$^{3+}$($O_h$) are antiferromagnetically coupled \cite{MAGilleo}.
Considering this and from the XRD, SAED and TEM results,
we conclude that the ferromagnetism in these nano-particles are intrinsic.
A schematic picture of hole-mediated exchange interaction between Fe$^{3+}$($O_h$) and Fe$^{2+}$($T_d$) ions is shown in Fig. 6.

%%%%%%%%%%%%%%%%%%%%%%
\section{Conclusion}
%%%%%%%%%%%%%%%%%%%%%%
In summary, we have investigated the electronic structure and magnetism of the
paramagnetic (Mn,Co)-co-doped ZnO and ferromagnetic (Fe,Co)-co-doped ZnO nano-particles using 2$p$$\rightarrow$3$d$ XAS and XMCD.
In the case of ZnO:(Mn,Co) nano-particles, the doped Mn and Co atoms are in a mixed-valence (2+, 3+, and 4+)
state and the relative concentrations of the high-valence (3+ and 4+) Mn and Co ions are higher in the surface region than in the deep core region.
Mn and Co 2$p$$\rightarrow$3$d$ XMCD results suggest that the paramagnetism comes from the Co$^{2+}$, Mn$^{3+}$ and Mn$^{4+}$ states.
In the case of the ZnO:(Fe,Co) nano-particles, too, the doped Fe and Co atoms
are found to be in a mixed-valence (2+ and 3+) state and
the relative concentrations of the Fe$^{3+}$ and Co$^{3+}$ ions are higher in the surface region than in the inner core region.
Fe and Co 2$p$$\rightarrow$3$d$ XMCD signals due to the ferromagnetic Fe ions and paramagnetic Fe and Co ions were observed in the surface region
while no appreciable XMCD signals were observed in the inner core region.
From these results, we suggest that the surface region is magnetically active and
Fe$^{3+}$ contributes to both the ferromagnetism and paramagnetism,
and that Fe$^{2+}$ contributes only to the ferromagnetism.
On the other hand, the ionic Co atoms in the surface region is paramagnetic and
that the ferromagnetic component of the Co ions is negligibly small.
Considering that the Fe$^{3+}$ ions are created due to Zn vacancies,
we conclude that the ferromagnetism of ZnO:(Fe,Co) nano-particles comes
from the hole-mediated exchange interaction between Fe$^{3+}$($O_h$) and Fe$^{2+}$($T_d$) in the surface region.

%%%%%%%%%%%%%%%%%%%%%%
\section{Acknowledgments}
The experiment at PF was approved by the Photon Factory Program Advisory Committee (Proposal No. 2008G010, 2010G187, and 2010S2-001).
The work was supported by a Grant-in-Aid for Scientific Research (S22224005) from JSPS, Japan,
a Global COE Program “the Physical Sciences Frontier", from MEXT, Japan,
an Indo-Japan Joint Research Project “Novel Magnetic Oxide Nano-Materials Investigated by Spectroscopy and ab-initio Theories" from JSPS, Japan, and
the Quantum Beam Technology Development Program Search and Development of Functional Materials Using Fast Polarization-Controlled Soft X-Rays from JST, Japan.

\section*{Figures}
\end{center}

\begin{figure}[htbp]
 \begin{center}
  \includegraphics[width=15cm]{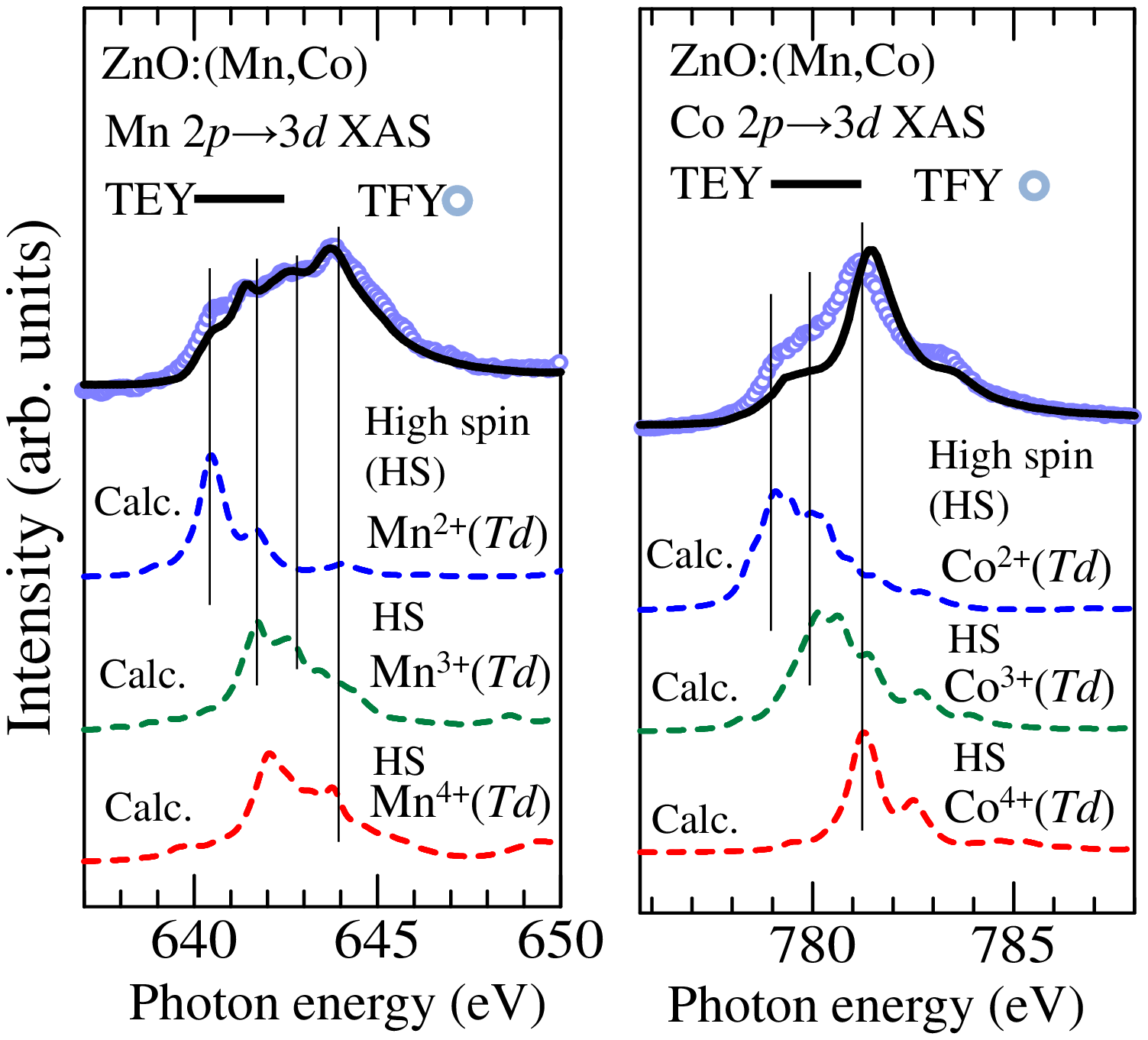}
 \end{center}
 \caption{(Color online)
Mn and Co 2$p$$\rightarrow$3$d$ XAS spectra (circles) of the paramagnetic ZnO:(Mn,Co) nano-particles taken both in the TEY and TFY modes.
Theoretical spectra (broken curves) of the Mn and Co ions in the 2+, 3+ and 4+ states calculated using the cluster model are shown.}
\end{figure}

\begin{figure}[htbp]
 \begin{center}
  \includegraphics[width=14cm]{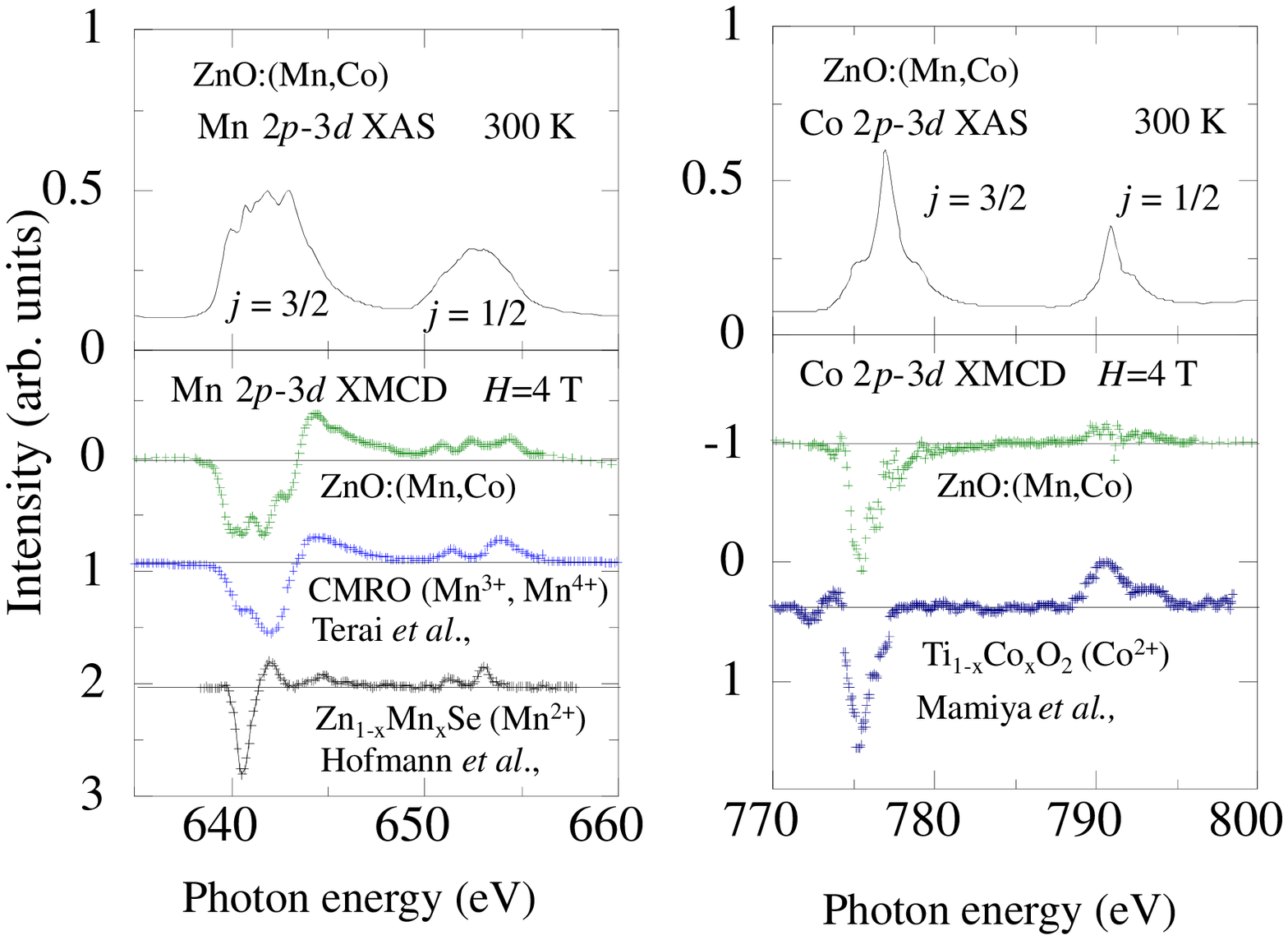}
 \end{center}
 \caption{(Color online)
Mn and Co 2$p$$\rightarrow$3$d$ XAS and XMCD spectra taken in the TEY mode in a magnetic field of $H$=4 T.
(a)Mn 2$p$$\rightarrow$3$d$ XAS and XMCD spectra of the paramagnetic ZnO:(Mn,Co) nanoparticles and
the Mn 2$p$$\rightarrow$3$d$ XMCD spectra of Ca$_{1-x}$Mn$_x$RuO \cite{Terai} and Zn$_{1-x}$Mn$_x$Se \cite{Hofmann}.
(b)Co 2$p$$\rightarrow$3$d$ XAS and XMCD spectra of the paramagnetic ZnO:(Mn,Co) nano-particles and
the Co 2$p$$\rightarrow$3$d$ XMCD spectrum of Ti$_{1-x}$Co$_x$O$_2$\cite{Mamiya}.}
\end{figure}

\begin{figure}[htbp]
 \begin{center}
  \includegraphics[width=12cm]{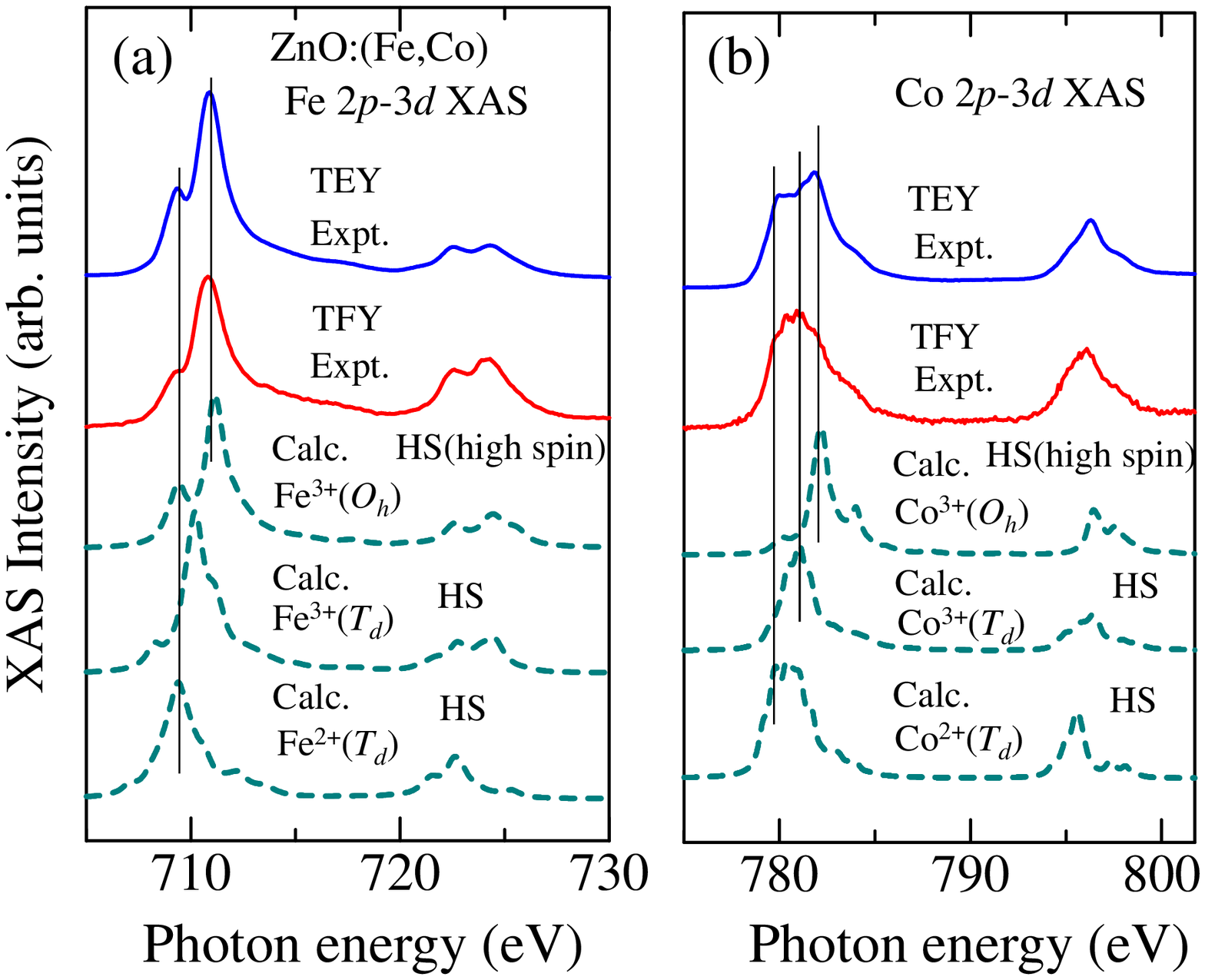}
 \end{center}
 \caption{(Color online)
Fe (a) and Co (b) 2$p$$\rightarrow$3$d$ XAS spectra of the ferromagnetic ZnO:(Fe,Co) nano-particles taken both in the TEY and TFY modes.
Theoretical spectra (broken curves) of the Fe and Co ions in the 2+ and 3+ states calculated using the CI cluster-model are shown by dashed curved.}
\end{figure}

\begin{figure}[htbp]
 \begin{center}
  \includegraphics[width=12cm]{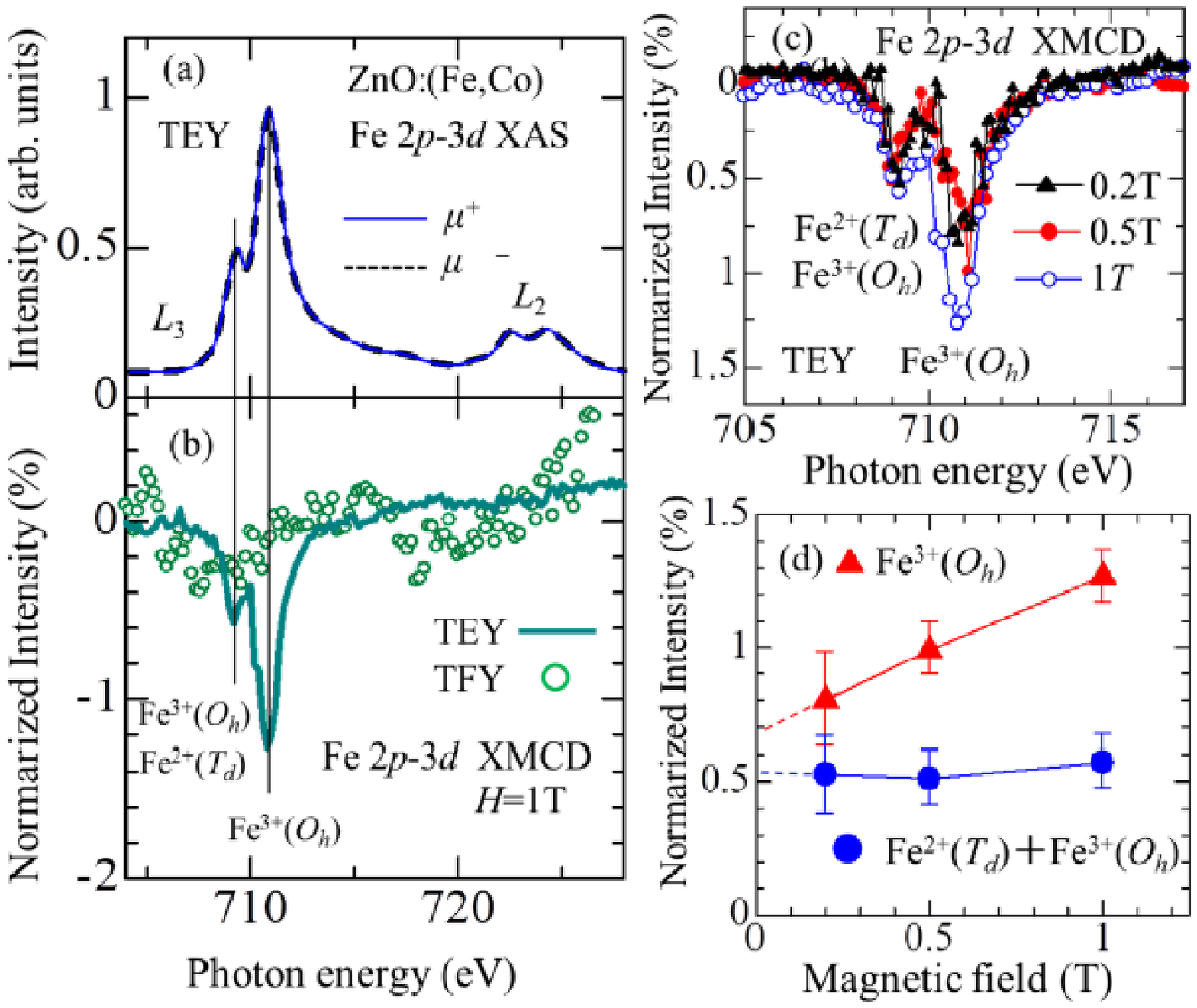}
 \end{center}
 \caption{(Color online)
Fe 2$p$$\rightarrow$3$d$ XAS and XMCD spectra of the ferromagnetic ZnO:(Fe,Co) nano-particles.
(a) Fe 2$p$$\rightarrow$3$d$ XAS spectra, taken in the TEY mode, in a magnetic field of $H$=1 T.
(b) Fe 2$p$$\rightarrow$3$d$ XMCD spectra taken both in the TEY and TFY modes.
(c) Fe 2$p$$\rightarrow$3$d$ XMCD spectra taken in the TEY mode at various magnetic fields.
(d) Fe 2$p$$\rightarrow$3$d$ XMCD intensity as a function of magnetic field.
}
\end{figure}

\begin{figure}[htbp]
 \begin{center}
  \includegraphics[width=12cm]{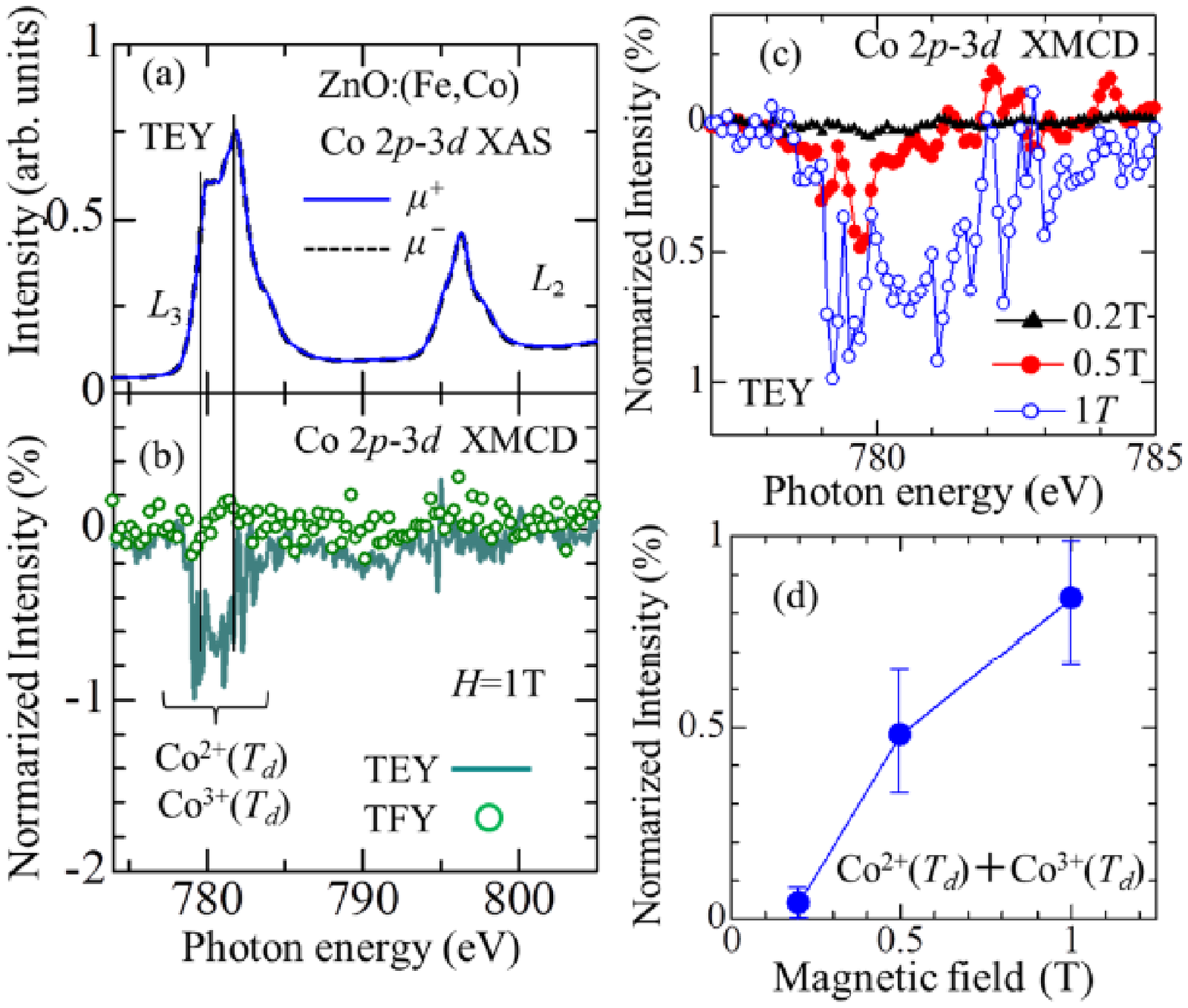}
 \end{center}
 \caption{(Color online)
Co 2$p$$\rightarrow$3$d$ XAS and XMCD spectra of the ferromagnetic ZnO:(Fe,Co) nano-particles.
(a) Co 2$p$$\rightarrow$3$d$ XAS spectra, taken in the TEY mode, in magnetic fields of $H$=1 T.
(b) Co 2$p$$\rightarrow$3$d$ XMCD spectra taken both in the TEY and TFY modes.
(c) Co 2$p$$\rightarrow$3$d$ XMCD spectra taken in the TEY mode at various magnetic fields.
(d) Co 2$p$$\rightarrow$3$d$ XMCD intensity as a function of magnetic field.
}
\end{figure}

\begin{figure}[htbp]
 \begin{center}
  \includegraphics[width=15cm]{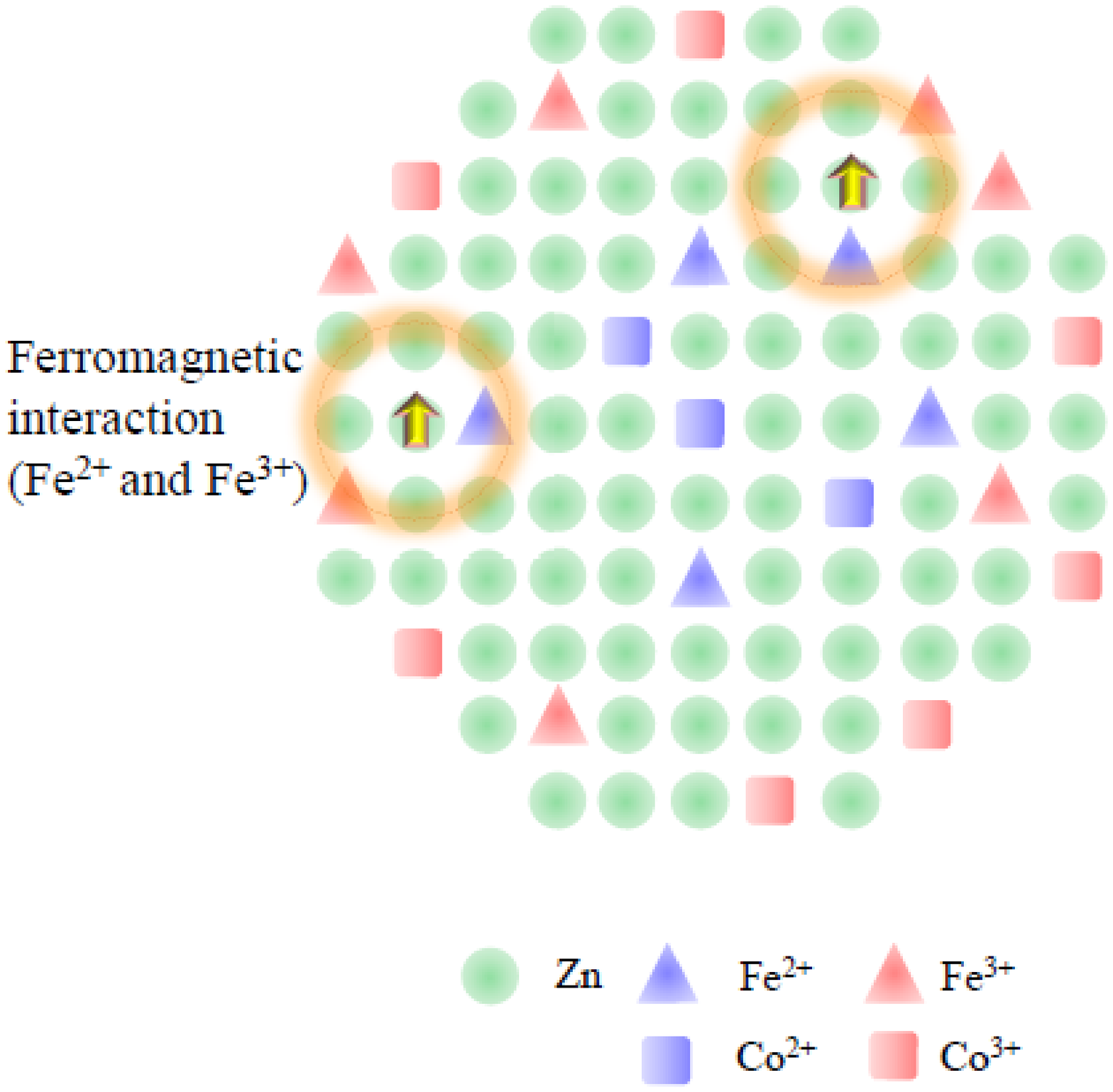}
 \end{center}
 \caption{(Color online)
A schematic of magnetic interactions in ZnO:(Fe,Co) nano-particles.
}
\end{figure}

\end{document}